\documentclass[11pt,a4paper]{article}
\usepackage[T1]{fontenc}
\usepackage[utf8]{inputenc}
\usepackage[margin=1in]{geometry}
\usepackage{graphicx}
\usepackage{booktabs}
\usepackage{tabularx}
\usepackage[hidelinks]{hyperref}
\makeatletter
\newcommand{\orcid}[1]{\g@addto@macro\@author{\\\href{https://orcid.org/#1}{ORCID: #1}}}
\newcommand{\affiliation}[1]{\g@addto@macro\@author{\\#1}}
\newcommand{\email}[1]{\g@addto@macro\@author{\\\href{mailto:#1}{#1}}}
\makeatother
\newcommand{\institution}[1]{#1\\}
\newcommand{\city}[1]{#1, }
\newcommand{\country}[1]{#1}
\newcommand{\socialsource}{\texorpdfstring{\textit{Social\textsuperscript{3} Source}}{Social3 Source}}
\title{\socialsource{}: Connecting Social Provenance,
Societal Qualities, and Social Purpose on an Open-Source Software Foundation}
\author{Jordi Cabot}
\orcid{0000-0003-2418-2489}
\affiliation{%
\institution{Luxembourg Institute of Science and Technology}
\institution{University of Luxembourg}
\city{Esch-sur-Alzette}
\country{Luxembourg}}
\email{jordi-cabot@list.lu}
\date{}
\begin{document}
\maketitle
\begin{abstract}
The speed at which software can be created and deployed today has increased dramatically.
But what is the point of all this software if it does not contribute to society?
The ultimate goal of building software should be bettering human lives and society.
And this goes beyond publishing the software as open source; it requires considering how it is developed, the societal qualities it embodies, and the purposes it serves.
This is what I call \socialsource{}.
This paper presents the concept, principles, and initial discussion for the implementation of \socialsource{}, illustrating how it integrates social considerations into open-source software development.
We have never generated as much software as we do today.
Let's make sure it counts.

\end{abstract}

\section{Introduction}

Creating software is easier than ever thanks to technologies such as low-code development~\cite{Cabot2024LowCode} and, especially, recent advances in artificial intelligence (AI).
These technologies help developers translate ideas into running applications more quickly, improving productivity and reducing the effort required to build software.\footnote{Whether this software meets appropriate standards of correctness, robustness, security... is a separate discussion.}
Their impact extends beyond professional developers: they also bring up (again!) the promise of \emph{citizen development}~\cite{Binzer2022Citizen}, and \emph{end-user development}\cite{Fischer2004MetaDesign}, in which non-technical people adapt and build software for their own needs.


This \emph{development fever}, however, comes at a sustainability cost.
As emphasized by the Karlskrona Manifesto, software sustainability involves interconnected environmental, economic, and social concerns, alongside individual and technical ones~\cite{Becker2015Sustainability}.
Creating, deploying, and operating more applications consumes resources, introduces maintenance and financial commitments, and can reproduce biases or reinforce exclusion.
Lowering the cost of creating an application does not eliminate the costs of sustaining it or the consequences of its use.
We should therefore ensure that the software we create also brings benefits to society, and assess those benefits alongside its costs and potential harms.
Paraphrasing Fei-Fei Li\footnote{\url{https://x.com/drfeifei/status/2102508817653379411}}:
\begin{quote}
The goal of building any software, AI included, should be bettering human lives and society.
\end{quote}

Open-source software grants essential freedoms to inspect, use, modify, and redistribute software~\cite{R3}.
These freedoms enable communities to understand, adapt, share, and maintain the applications on which they depend.
We take open source as a necessary foundation for our proposal, but it is not a sufficient condition to guarantee that software benefits society.
An open license alone does not establish that development was inclusive, that an application embodies relevant societal qualities, or that its use benefits a relevant community.

To connect these concerns on top of the open source foundation, we introduce \socialsource{}: open-source software that is social in how it is built, in the qualities it embodies, and in the purposes it serves.
The superscript three refers to three complementary social dimensions.
\emph{Social provenance} concerns how and by whom software is developed and governed, including diverse contributions and meaningful participation by all the affected communities.
\emph{Societal qualities} concern the properties the software embodies, such as accessibility, multilingualism, privacy, fairness, affordability, and environmental sustainability.
\emph{Social purpose} concerns the benefitial impact expected from the use of that software.

\begin{center}
\setlength{\fboxsep}{10pt}
\fbox{%
\begin{minipage}{\dimexpr\linewidth-2\fboxsep-2\fboxrule\relax}
\small
\textbf{An illustrative example: an application for migrant services}
\par\medskip
Consider a community organization developing an open-source application to help migrants find local services and navigate administrative procedures.
Its open license would allow other organizations to inspect, adapt, redistribute, and maintain it.
The scenario illustrates how the three social dimensions of \socialsource{} could be put into practice and assessed.

\medskip
\textbf{Social provenance.}
Migrants, social workers, translators, accessibility specialists, and developers would jointly define requirements, test the application, and influence project decisions.
Documenting their contributions and how their feedback changes the application would provide evidence of meaningful participation beyond code contributions.

\medskip
\textbf{Societal qualities.}
The application would offer appropriate languages, accessible interaction, minimal collection of sensitive information, and support for modest devices and limited connectivity.
Accessibility evaluations, privacy checks, and testing with intended users would help establish whether these qualities are achieved in practice.
Trade-offs would require explicit decisions: for example, AI translation could broaden language coverage while introducing translation errors, resource costs, and reliance on an internet connection.

\medskip
\textbf{Social purpose.}
The application would address service-access barriers identified by the community.
Evaluation would examine whether migrants can find relevant services and complete administrative tasks more easily thanks to the app, while also investigating unintended harms and who remains excluded.
These observations would provide evidence of societal benefit beyond a stated intention to help.

\end{minipage}%
}
\end{center}

The remainder of the paper is structured as follows.
Section~\ref{sec:concept} defines the \socialsource{} concept in more detail.
Sections~\ref{sec:provenance}--\ref{sec:purpose} describes the three social dimensions.
Section~\ref{sec:besser} presents a proposed development framework to support the concept, and Section~\ref{sec:related} discusses related concepts and initiatives.
Section~\ref{sec:conclusion} concludes the paper and presents a roadmap of research challenges.

\section{The \socialsource{} concept}
\label{sec:concept}

\begin{figure}[htbp]
\centering
\includegraphics[width=0.9\linewidth]{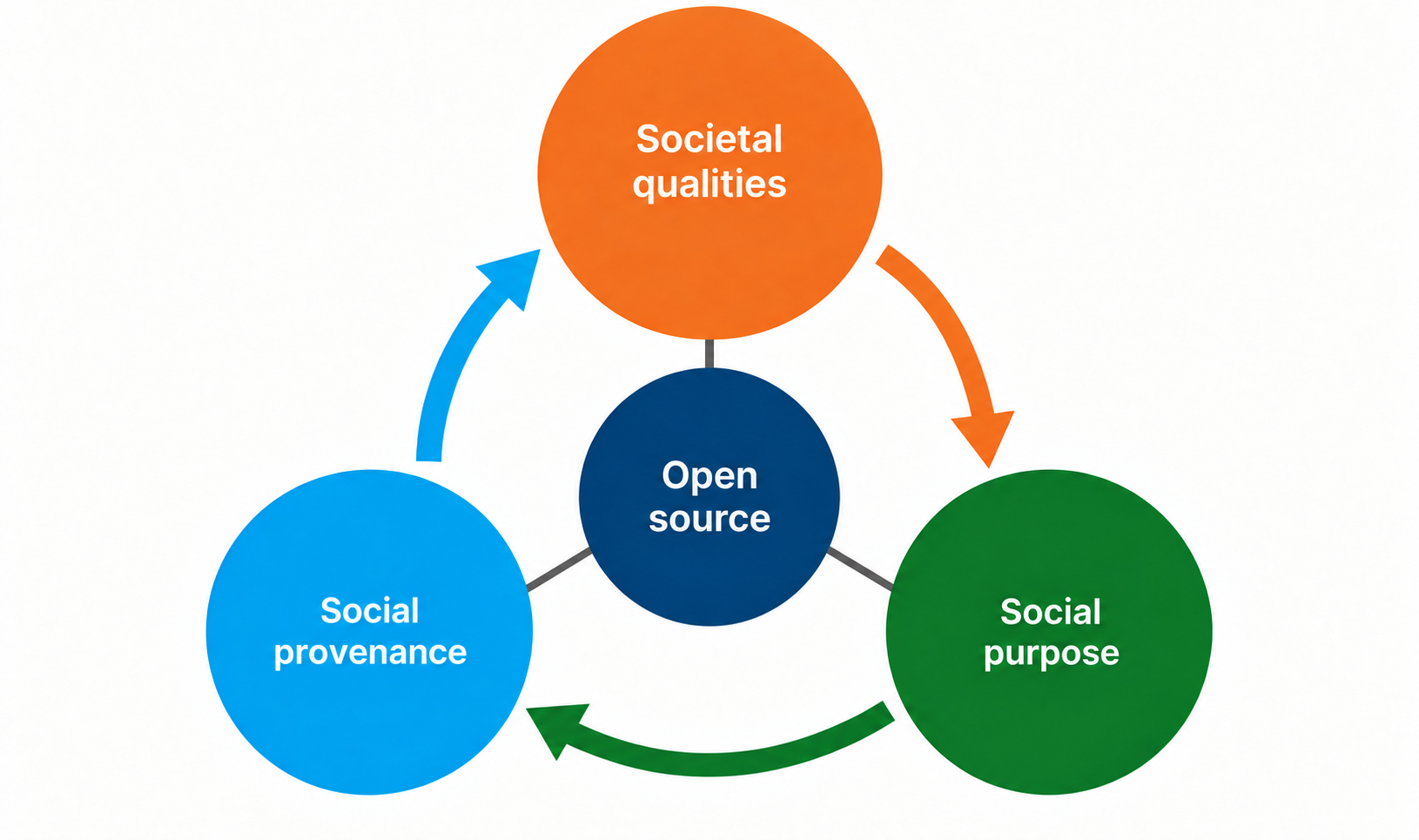}
\caption{The three social dimensions of \socialsource{} share an open-source foundation.
The surrounding arrows emphasize their interdependence across software development, deployment, and evolution.}
\label{fig:socialsource-foundation}
\end{figure}

We define \socialsource{} as \emph{open-source software that is social in how it is built, in the qualities it embodies, and in the purposes it serves}.
As illustrated in Figure~\ref{fig:socialsource-foundation}, open source provides their shared foundation, while the connections among the dimensions emphasize that decisions about people, software properties, and intended benefits influence one another.

The open-source foundation establishes the freedoms to inspect, use, modify, and redistribute the software~\cite{R3}.
These freedoms allow communities to adapt an application to their circumstances, share improvements, and continue its development beyond the involvement of its original creators esuring its long-term survival \cite{survivalrate} chances.
Open governance, accessible documentation, and open standards can help communities exercise these freedoms.
Within our proposal, all \socialsource{} software is therefore open source, while an open-source license alone does not establish the three social dimensions.

\textbf{Social provenance: how and by whom the software is built.}
This dimension concerns the people (going far beyond the developers), contributions, and decisions that shape an application throughout its development and maintenance.
The aim is to enable people with different backgrounds and relevant experience, particularly affected communities, to influence requirements and decisions through inclusive contribution routes, transparent governance, and fair recognition.

\textbf{Societal qualities: what properties the software embodies.}
This dimension concerns how an application's design and operation reflect the needs, resources, and circumstances of the people who use it and makes sure they are all ablee to effectively use and benefit from the software. 
Relevant qualities include accessibility, multilingualism, affordability, privacy, fairness, security, and environmental sustainability.
Their priorities and the trade-offs among them depend on the context and should be expressed as requirements that can guide implementation and evaluation.

\textbf{Social purpose: why the software exists and whom it serves.}
This dimension concerns the social needs an application addresses and the consequences of its use for its intended communities.
Software may acquire such a purpose through its original design or through later adaptation and adoption by a community.

Addressing even one of these dimensions is already valuable, and efforts that advance one or two dimensions deserve recognition even when they do not yet address all three.
The aim of \socialsource{} is to encourage projects to think about all three and adopt them as early possible in the development process and the whole software lifecycle. 

\socialsource{} applies regardless of the development technology or whether a project operates commercially, within a non-governmental organization (NGO), or in another organizational setting.

Achieving these properties also involves trade-offs.
Unfortunately, developing \socialsource{} software will often require more time and a larger budget than delivering the same core functionality without these commitments.
Involving more people during conception and development requires time for participation, coordination, and collective decision-making.
Multilingual support and accessibility may require additional features, design work, and validation with users.
Reducing bias in AI-enabled functionality may also require selecting, adapting, or training models for the intended communities, together with appropriate data collection and evaluation.
These efforts introduce costs that need to be anticipated when defining the project's scope, resources, and priorities.
They also illustrate why the pursuit of societal qualities requires explicit choices about acceptable compromises and minimum requirements.

\section{Social provenance}
\label{sec:provenance}

Recognizing the importance of non-coding roles is a first step toward diversity in software development ~\cite{Canovas2022NonCoding}. 
However, recognizing a variety of roles is not enough.
People in both coding and non-coding roles should be able to actively participate and influence the project direction. 
Moreover, these roles should also be occupied by people with diverse backgrounds, origins, and lived experiences, with the relevant forms of diversity informed by the software's context and domain, intended user communities, and anticipated usage scenarios.
This diverse provenance and expertise maximizes the chances of the final \socialsource{} application to meet its goals.

\begin{figure}[!htbp]
\centering
\includegraphics[width=0.68\linewidth,height=0.32\textheight,keepaspectratio]{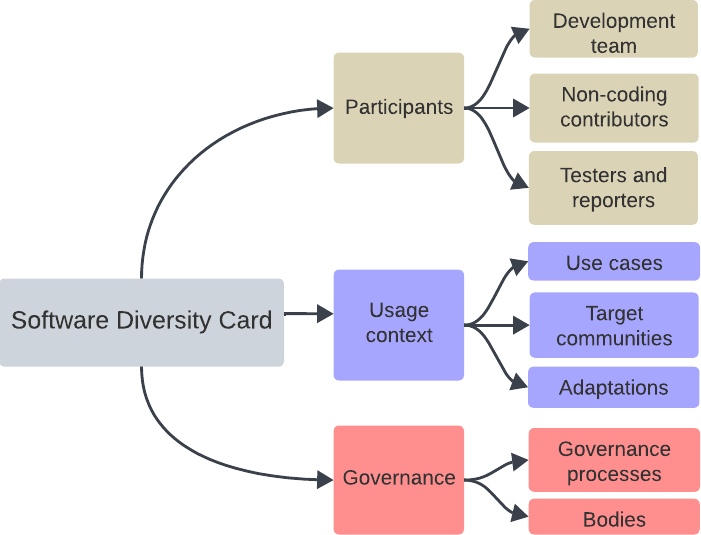}
\caption{Main components of a Software Diversity Card, reproduced from ~\cite{R2}.}
\label{fig:diversity-card}
\end{figure}

Social provenance, therefore, concerns both the range of contributions a project welcomes and whose perspectives and influence those contributions represent. Following the proposal from the Software Diversity Card~\cite{R2}, we decompose the provenance dimension of software into \textit{participants}, \textit{usage context}, and \textit{governance}.
Figure~\ref{fig:diversity-card} depicts these three subdimensions.
Participants describe those roles and individuals contributing within a project, usage context describes the communities and circumstances it serves, and governance connects participants and users via a set of rules and policies to make community decisions.

\textbf{Participants.}
Participants include development teams, translators, documenters, issue reporters, and users involved in testing or feedback.
Relevant attributes include their roles, languages, experience, education, location, and demographic characteristics, where appropriate to collect and disclose.
This broader view matters because repository commits capture only part of the work, and people who report issues or join tests may not represent the whole set of intended users.
In the migrant-services example, provenance would therefore document the involvement of migrants, social workers, and translators alongside developers, including how their feedback changes requirements.

\textbf{Usage context.}
Usage context encompasses intended use cases, target communities, social and cultural circumstances, and adaptations for particular groups.
Relevant characteristics include language, digital literacy, accessibility needs, and the resources available to users.
For the running example, language and digital-literacy differences would inform some of the adapations and qualities the software should include and that would need to be therefore evaluated.

\textbf{Governance.}
Governance encompasses the bodies, decision-makers, and rules that direct a project, together with its organizational setting, funders, and shareholders.
For \socialsource{}, this information should help explain who (and how and when) prioritizes requirements, accepts contributions, and resolves disagreements.
A diverse contributor population has limited influence if decisions remain inaccessible to those contributors.
Documenting governance \cite{GovernanceDSL}  can therefore support scrutiny of whether community participation affects the application's direction.

\section{Societal qualities}
\label{sec:qualities}

Software must embed societal qualities that help realize and respect human and social values. Examples of well-known properties expected from a a social software are: Accessibility, multilingualism, multiculturalism, privacy, fairness, environmental sustainability, etc. 

But they are not the only ones. Work in the social sciences can provide a more top-down approach to identifying relevant properties, starting from general frameworks of human and social values and translating them into software qualities suited to people's needs and circumstances. We propose to start by drawing on two complementary perspectives. 

On the one hand, Schwartz's Theory of Basic Human Values describes 19 values, covering concerns such as equality and justice, acceptance of difference, protection of nature, and autonomy of thought and action~\cite{Schwartz2012Refining}.

We could then refine these values in more concrete software qualities and requirements. Table~\ref{tab:values-qualities} illustrates shows a partial mapping for some values.

\begin{table}[htbp]
\centering
\caption{Illustrative mappings from human values to candidate software qualities for \socialsource{}.}
\label{tab:values-qualities}
\small
\renewcommand{\arraystretch}{1.2}
\begin{tabularx}{\linewidth}{@{}>{\raggedright\arraybackslash}p{0.34\linewidth}>{\raggedright\arraybackslash}X@{}}
\toprule
\textbf{Human value} & \textbf{Candidate software qualities or requirements} \\
\midrule
Universalism--concern & Accessibility, fairness, and equitable access \\
Universalism--tolerance & Multilingualism and respect for cultural differences \\
Universalism--nature & Energy efficiency and reduced environmental impact \\
Self-direction--thought/action & User autonomy, understandable choices, and adaptability \\
Security--personal & Privacy, data protection, and protection against harm \\
Benevolence--caring & Inclusion for all members of the community being served \\
\bottomrule
\end{tabularx}
\end{table}

Note that these relationships are neither exhaustive nor one-to-one.
Privacy can support both security and autonomy, while equitable access may require accessibility, affordability, and support for limited (internet) connectivity.
And, as always, ideally we would all software satisfy all qualities but technical, economical, temporal,.. trade-offs are unavoidable when deciding which ones to prioritize and actually implement. 

On the other hand, the Capability Approach of Sen and Nussbaum complements the values discussion by stating that values are not enough and we need to make sure that people have actual opportunities to pursue those valued  activities~\cite{Nussbaum2003Capabilities}.
Indeed, implementing a desirable software property does not ensure that users can benefit from it.
For example, multilingual support may not bring any value if the users that would need that support cannot activate it due to a poor UI design or lack of connectivity (imagine English support is local but any other language requires an on-the-fly AI-based translation that requires internet access).

\section{Social purpose}
\label{sec:purpose}

The most important aspect of \socialsource{} is that its code implements something useful for our society.

It's true that, in principle, almost any software could fit this definition by helping an organization with a social purpose to perform any of its digital tasks. An NGO, for example, may use a spreadsheet, an accounting package, or a communication platform to organize its work.
This potential usefulness is valuable, but it provides little basis for distinguishing software with a \textit{true} social purpose: under such a broad interpretation, practically any application could qualify.

In practice, organizations pursuing social missions often need adaptations to common software because their structures, resources, and ways of working differ from those assumed by standard products.
A contact-management application, for instance, may need to represent specific roles (beneficiaries, volunteers,..) and offer specific features (e.g. protection identities for volunteers that may suffer state violence if recognized)

More importantly, these organizations need software that supports their core mission-oriented activities.
Examples could include coordinating food distribution, matching volunteers with requests for assistance, managing access to shelters, or helping migrants navigate local services and administrative procedures as in our illustrative scenario.


This is what we mean by software with a \emph{social purpose}: software created or adapted to help organizations and communities pursuing social missions do their work better.
The connection to the mission should be explicit, identifying whose work or experience the software is intended to improve, which activities it supports, and what barriers it seeks to reduce.
NGOs are an important example of the type of organization this software should help, but the same reasoning applies to public services, cooperatives, informal community groups, and other organizations pursuing social goals.

\section{A model-driven development framework for \socialsource{}}
\label{sec:besser}

The \socialsource{} conceptual framework should be actionable in order to guarantee its adoption.

To facilitate its integration into the development process alongside the application's functional requirements, I propose a model-driveen approach for \socialsource{}.

Just as models can specify an application's data, behavior, and interfaces and drive its generation, they could also describe the community around the software, the societal qualities it is expected to exhibit and its overarching goals.
These specifications would make social commitments explicit, connect them to implementation decisions, and support their review as the application evolves.

This approach could be implemented by extending existing modeling and low-code platforms such as BESSER~\cite{Alfonso2024BESSER}\footnote{\url{https://besser-pearl.org/}}.
BESSER provides an open-source foundation of modeling facilities and extensible generators on which additional social specifications could be built extending existing modeling languages (from goal models \cite{goalmodeling} to governance DSLs \cite{GovernanceDSL}) or creating new ones for this purpose. 
Then, graphical editors, forms, and natural-language interaction could allow members with different technical backgrounds to contribute to these specifications and inspect the resulting models before code is generated.

\begin{figure}[htbp]
\centering
\includegraphics[width=\linewidth]{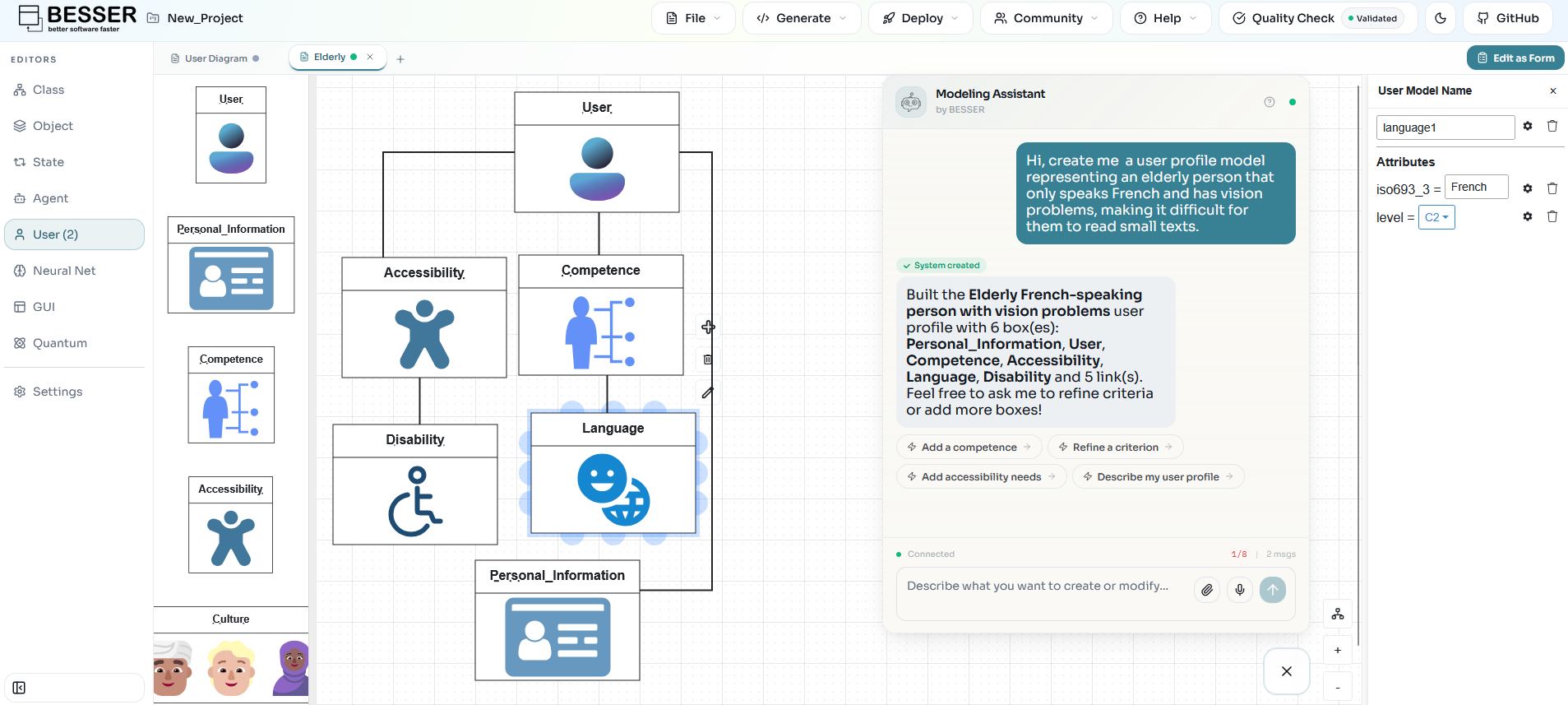}
\caption{User-profile modeling in BESSER with a conversational assistant based on ~\cite{Conrardy2026Personalization}.}
\label{fig:besser-user-profiles}
\end{figure}

As an example, Figure~\ref{fig:besser-user-profiles} illustrates a starting point for modeling the people around an application, a key element in the provenance dimension. The screenshot shows a user profile model in the BESSER editor describing that we can have users with accessibility needs and different language competences in the application. Next to the graphical canvas, a conversational assistant helps constructing the editable graphical user profiling model. 
As we said above, other types of models would be needed to cover all the other social aspects of \socialsource{}.

This information could then be either stored in the software repository in a machine-readable format (e.g. a \texttt{governance.md} file for the governance policies) or could influence the code generators to make sure the generated code embeds the required software qualities. 
Indeed, in our model-driven \socialsource{} approach, generators should go beyond rule-based templates only focused on implementing the functional requirements and become more flexible to accommodate social requirements.

\section{Related work}
\label{sec:related}

The idea that software can contribute to society and the public good is well established.
Ferrario et al.~\cite{Ferrario2014SocialGood} explicitly advocate software engineering for positive social change and propose Speedplay, a framework combining agile development, action research, and participatory design.
Whittle et al.~\cite{Whittle2021HumanValues} further argue that human values need to be integrated into the software engineering method itself, rather than considered only as properties of the final software.
Their work connects social purpose with development practices that account for the communities involved.
Abebe et al.~\cite{Abebe2020SocialChange} examine the roles computing can play in social change while emphasizing the limitations of technical interventions: increasing our capacity to produce software does not, by itself, resolve social problems.
Humanitarian free and open-source software (HFOSS) is also an established precedent for open software intended to benefit people.
Ellis et al.~\cite{Ellis2015HFOSS} study students' participation in HFOSS projects, providing evidence about educational experiences rather than measured benefits for recipient communities.

Other work addresses the specification and modeling of speecific societal qualities.
Moreira et al.~\cite{Moreira2023Sustainability} propose a catalogue of social and technical sustainability requirements, represented through feature models and iStar, including concerns such as fairness.
Their catalogue provides a precedent for translating broad concerns into reusable requirements and a potential input to the societal subdimensions and requirements DSLs envisaged for \socialsource{}.
Roy et al.~\cite{Roy2026Taxonomy} organize sustainability requirements across environmental, social, technical, and economic dimensions, discussing metrics and relationships between categories. 
These relationships are relevant to identifying compatible requirements and negotiating acceptable trade-offs.
Our LangBiTe approach~\cite{Morales2024LangBiTe} offers a more focused example: a DSL for specifying ethical requirements and generating tests for fairness and bias in large language models, with traceability between requirements and their assessment.
It illustrates how a particular societal concern can be integrated into a model-driven development process.

For social provenance, support for explicitly modeling and reporting the diversity of contributors remains limited.
The Software Diversity Card~\cite{R2}, introduced earlier, is a direct foundation for describing this aspect.
Related work focused on user modeling addresses a complementary concern: representing the people expected to use an application.
Our systematic literature review~\cite{Conrardy2025UserModeling} identifies fragmented approaches to user modeling in model-driven engineering, with limited coverage of user characteristics and tool support.
Such models can inform the profiling of community members, but describing users alone does not establish who participates in development, whose perspectives are represented, or who has decision-making authority.

These contributions provide some of the building blocks for \socialsource{}. 
We extend them and bring them together as part of a unified and end-to-end process
making the framework actionable. 

\socialsource{} also complements initiatives such as the Digital Public Goods Alliance~\cite{DPGA}, which promotes the discovery and adoption of digital public goods, including software serving social needs, through its registry, advocacy, and guidance on best practices.
Our focus on integrating social concerns into software engineering methods could support these efforts by helping teams translate such aims and practices into explicit models, development decisions, and implementations.

\section{Conclusions and roadmap}
\label{sec:discussion}
\label{sec:conclusion}

This paper introduced \socialsource{}, open-source software that is social in how it is built, in the qualities it embodies, and in the purposes it serves.
We also proposed integrating these dimensions into model-based development by extending modeling platforms with new social models, explicit societal requirements, and generators that act on those requirements.

The resulting framework is an initial position for discussion; realizing and evaluating it requires progress on many research challenges including the ones highlighted in what follows.

\subsection{Strengthening interdisciplinary foundations}

Developing \socialsource{} requires closer collaboration with researchers in the social sciences, economics, and political science to align the approach with, and learn from, their theories, methods, and empirical findings.
A key objective is to build a comprehensive taxonomy and ontology of the human and social values that software qualities should help fulfill, extending the initial perspectives discussed in Section~\ref{sec:qualities}.
This work should clarify relationships and tensions between values and connect them to concrete software qualities, contextual requirements, and evidence and scoring of their fulfillment.

\subsection{Defining metrics and a social score}
Linked to the above challenge, a core difficulty is to define metrics for each social dimension and use them to produce a meaningful \socialsource{} score that could then be used to compare software products targeting the same functionality. 
Research is needed to select and validate indicators, normalize measurements, and determine how priorities should influence their weighting.
An overall score should also report the evidence used to compute the score, its completeness and the uncertainty scoring for some dimensions (e.g. linked to the use of AI components for the implementation of some social aspects and the inherent non-determinism that comes with it).
Qualification thresholds and comparisons must also account for differences in project scale, maturity, deployment context, and available resources.

\subsection{A complete set of social modeling  DSLs and generators}

The framework requires domain-specific languages (DSLs) that can express all three social dimensions and connect them to the application's functional models.
Existing languages for user profiles, diversity, governance, and software requirements provide starting points, but they do not yet suffice. 
A combination of language extensions and new DSLs should fill the remaining gaps and distinguish mandatory constraints from negotiable preferences, while recording dependencies, priorities, and decision rationales.
To leverage all this information and guarantee the ROI of creating such models, a brand new set of code-generators should be able to use them as input to generate new software artefacts (e.g. for documentation) and to influence the creation of others (e.g. to automatically add internationalization features when required). 

\subsection{Building a library of socially aware generators and architectures}

A reusable library of socially aware generators, templates, and reference architectures could transform social requirements into social implementations without having to start from scratch every single time.
The longer-term objective is to support certified generators linked to explicitly defined social requirements, with each reusable element specifying its assumptions, supported properties, and verification evidence.

\subsection{Negotiating priorities and acceptable trade-offs}

Expressing requirements does not resolve disagreements about which qualities matter most or whose preferences should prevail.
We must develop participatory modeling methods for eliciting priorities, representing affected communities, resolving conflicts, and revisiting explicit decisions as circumstances change.
For example, a community may prefer locally maintained translations over an external AI service for sensitive material, while accepting narrower language coverage.
Traceability is also needed to link each trade-off and its rationale to the affected requirements, implementation choices, and evidence collected during operation.

\subsection{Making development tools and processes accessible}

Accessibility must extend to the tools and processes used to create the software not only be aprt of the resulting application.
Following the Capability Approach~\cite{Nussbaum2003Capabilities}, an invitation to participate is insufficient unless team members have effective opportunities to express requirements, inspect and modify models, test alternatives, and influence decisions.
Modeling environments, testing approaches, documentation, and collaboration practices must therefore accommodate participants with different abilities, languages, resources, and levels of technical knowledge.
One research direction is to extend these tools with AI-powered interfaces, including voice interaction, conversational assistants, and augmented reality, offering alternative ways to create, explore, and validate models.
Such interfaces could help non-technical participants contribute domain knowledge without first mastering a modeling language.

\subsection{Validating usefulness and social benefit in practice}

The framework, its metrics, and its supporting tools require evaluation with organizations and communities pursuing different social missions.
Case studies should examine whether the approach helps participants express their needs, improves development decisions, and leads to useful software at an acceptable cost.
These evaluations should also test whether the proposed dimensions are sufficient and refine the framework in response to the people it aims to serve.

\nocite{*}
\bibliographystyle{unsrt}
\bibliography{references}
\end{document}